\documentclass[preprintnumbers,amsmath,amssymb,nofootinbib,superscriptaddress,floatfix,showkeys,twocolumn]{revtex4-1}
\usepackage{color,amsmath,amssymb,graphicx,latexsym,subfigure}
\usepackage{threeparttable,txfonts}
\usepackage{appendix}
\usepackage{ulem}
\usepackage{float}
\usepackage{graphicx}

\newcommand{\gev}{\ensuremath{\,\mathrm{GeV}}}
\newcommand{\tev}{\ensuremath{\,\mathrm{TeV}}}

\newcommand{\sigsip}{\ensuremath{\sigma^{\rm{SI}}_{\chi p}  }}

\newcommand{\sigsdn}{\ensuremath{\sigma^{\rm{SD}}_{\chi n}  }}
\newcommand{\sv}{\ensuremath{\langle\sigma v\rangle}}
\newcommand{\gtwo}{\ensuremath{\delta a_\mu}}

\begin{document}

\title{A common origin of muon g-2 anomaly, Galaxy Center GeV excess and AMS-02 anti-proton excess in the NMSSM}

\author{Murat Abdughani$^{a}$}
\author{Yi-Zhong Fan$^{a,b}$}
\email{yzfan@pmo.ac.cn}
\author{Lei Feng$^{a,b,c}$}
\author{Yue-Lin Sming Tsai$^a$}
\email{smingtsai@pmo.ac.cn}
\author{Lei Wu$^d$}
\email{leiwu@njnu.edu.cn}
\author{Qiang Yuan$^{a,b}$}

\affiliation{
$^a$Key Laboratory of Dark Matter and Space Astronomy, Purple
Mountain Observatory, Chinese Academy of Sciences, Nanjing
210023, P.R.China \\
$^b$School of Astronomy and Space Science, University of Science and Technology of China, 
Hefei, Anhui 230026, China\\
$^c$Joint Center for Particle, Nuclear Physics and Cosmology,  
Nanjing University -- Purple Mountain Observatory,  Nanjing  210093, China \\
$^d$Department of Physics and Institute of Theoretical Physics, 
Nanjing Normal University, Nanjing, 210023, China 
}

\begin{abstract}
The supersymmetric model is one of the most attractive extensions of the Standard Model of particle physics. 
In light of the most recently reported anomaly of the muon g-2 measurement by the FermiLab E989 experiment, 
and the excesses of gamma rays at the Galactic center observed by Fermi-LAT space telescope, 
as well as the antiproton excess observed by the Alpha Magnetic Spectrometer, 
we propose to account for all these anomalies or excesses in the Next-to-Minimal Supersymmetric Standard Model. 
Considering various experimental constraints including the Higgs mass, B-physics, 
collider data, dark matter relic density and direct detections, 
we find that a $\sim 60$ GeV bino-like neutralino is able to successfully explain all these observations. Our scenario can be sensitively probed by future direct detection experiments.

\end{abstract}
\date{\today}

\maketitle

\section{Introduction}

The muon anomalous magnetic moment $a_\mu$ has been very recently measured by  
E989 at Fermilab with an unprecedentedly relative precision of 368 parts-per-billion (ppb).  
By combining the new data with the previous measurement from Brookhaven National Lab (BNL)~\cite{Tanabashi:2018oca}, 
they found a deviation $\delta a_\mu= (2.51 \pm 0.59) \times 10^{-9}$ 
with $4.2\sigma$ significance~\cite{2104.03281} from
the Standard Model (SM) prediction.
This indeed calls for the new physics beyond the Standard Model (BSM) \cite{hep-ph/0102122, 1610.06587,Aoyama:2020ynm,1704.02078,2003.03386,2008.02377,1505.05877,1606.05329,1906.08768,2012.03928,1908.00921}. 

Meanwhile, various astrophysical and cosmological observations 
show that the dark matter (DM) constitutes the majority of matter in the universe. 
Among many DM candidates, the Weakly Interacting Massive Particles (WIMPs) 
have been a compelling candidate. 
The appeal of WIMP DM is due in part to the suggestive coincidence between the thermal abundance of WIMPs 
and the observed dark matter density through the thermal freeze-out mechanism, known as the \textit{WIMP miracle}. 
There have been various experiments devoted in the quest of nature of DM. 
The indirect detections of DM from searching for the gamma rays and cosmic rays 
have reported some intriguing excesses. For examples, 
the Galaxy center GeV gamma-ray excess from the Fermi LAT (i.e., the so-called GCE, 
see~\cite{1010.2752,1402.6703,1406.6948,Calore:2014xka,1511.02621,1704.03910,1812.06986}) 
and the possible anti-proton excess from AMS-02 collaborations~\cite{1610.03840,1610.03071,1903.02549,1903.01472,PhysRevResearch.2.043017},  
they can be consistently interpreted as the $\sim 50-100$ GeV dark matter annihilating 
into the $b\bar{b}$ final states.

In conjuncture with the muon g-2 anomaly, 
all these anomalies may indicate the new physics in dark sector. 
Supersymmetry (SUSY) naturally provides the DM candidate, 
such as the lightest neutralino for $R$-parity conserving scenario. 
In SUSY, the muon $g-2$ anomaly can be explained by the contributions of light electroweakinos and sleptons 
running in the loops~\cite{Moroi:1995yh,Fowlie:2013oua,Martin:2001st,Stockinger:2006zn,1610.06587,Abe:2002eq,hep-ph/0102122,1909.07792}. 
Although the low mass neutralino in the CP-conserving Minimal Supersymmetric Standard Model 
is still favored by muon $g-2$ anomaly, it cannot interpret the GCE because of $p$-wave 
suppressed annihilation cross section~\cite{Carena:2019pwq}.  
On the other hand, a new CP-odd singlet in the Next-to-Minimal Supersymmetric Standard Model (NMSSM) 
can play the role of a mediator in DM annihilation~\cite{Kowalska:2012gs}. 
Once the mass of this singlet Higgs is just as heavy as two neutralino masses (singlet Higgs resonance), 
neutralino can effectively $s$-wave annihilate to the $b \bar{b}$ final state at the zero temperature.  
Consequently, the parameter space allowed by muon g-2 measurement may coincide 
with the GCE and anti-proton excess. 

In this Letter, we first perform a state-of-art analysis and find out that
the anomalous muon $g-2$, GCE and the anti-proton excess may have 
a common physical origin that relates with DM in the NMSSM. 
We then show that  such a scenario can be effectively probed in the future DM direct detection (DD) experiments.

\section{Model and Methodology}
In the scale invariant NMSSM~\cite{Konig:1991tr}, 
a $Z_3$ symmetric gauge singlet chiral superfield $\hat S$ is introduced.
In addition to MSSM, the superpotential is
\begin{equation}
    W = W_\mathrm{MSSM} + \lambda S H_u H_d + \frac{\kappa}{3} S^3,
\end{equation}
where the new singlet Higgs develops a vev $\langle S\rangle=s$. 
The superpartner of $S$ (singlino) can mix with gaugino and Higgsino 
as the neutralino mass matrix 
\begin{eqnarray}
M_{\chi^0} &=& \left( \begin{array}{ccccc}
   M_1 & 0 & -m_Z c_\beta s_W & m_Z s_\beta s_W  & 0 \\
   &M_2& m_Z c_\beta c_W & -m_Z s_\beta c_W  & 0 \\
	      &   &                  0 & -\mu    & -\lambda v_u\\
	      &  &   &  0 & -\lambda v_d \\
	      &  &   &   & \frac{2\kappa}{\lambda} \mu
	\end{array} \right).
	\label{neutralinomassmatrix}
\end{eqnarray}
The $5 \times 5$ unitary matrix is defined in the group basis  
(Bino $B^0$, Wino $W^0$, Higgsino $h_{u}$, Higgsino $h_{d}$, Singlino $s^0$). 
The effective $\mu$-term is defined by $\lambda s$ and  
the $Z$-boson mass is $m_Z$.   
The vacuum expectation values for $h_u$ and $h_d$ are denoted as $v_u$ and $v_d$. 
Their ratio is $\tan \beta = v_u / v_d$ and we define $s_\beta=\sin\beta$ and $c_\beta=\cos\beta$.  
Similarly, the sine and cosine of Weinberg angle are $s_W$ and $c_W$.
The gaugino mass $M_1$ and $M_2$ are the soft bino and wino masses. 
After diagonalized, the lightest neutralino $\chi^0_1$ can be DM by assuming $R-$parity conserved.

In this Letter, we narrow down the list of the NMSSM parameters to nine free inputs 
and Their prior ranges are 
\begin{equation}
\begin{aligned} 
 &0.001 < \lambda < 1,~0.001 < |\kappa| < 2,~ |A_\lambda| < 3000,~ |A_\kappa| < 20, \\
 &30 ~\mathrm{GeV} < M_1 < 80 ~\mathrm{GeV},~100 ~\mathrm{GeV} < M_2 < 1000 ~\mathrm{GeV}, \\
 &100 ~\mathrm{GeV} < |\mu| < 1000 ~\mathrm{GeV},~100 ~\mathrm{GeV} < M_{\tilde \ell_{1,2}} < 1000 ~\mathrm{GeV}, \\
 &1 < \tan\beta < 60, 
\label{eq:scan}
\end{aligned}
\end{equation}
where the soft-breaking mass parameters of the electroweakinos and sleptons 
are chosen to be less than $1\tev$ to produce a sizable positive corrections to the muon $g-2$~\cite{0806.0733}, 
while the small $M_1$ is required to produce the GCE spectra~\cite{Calore:2014xka}. 
Since our studied observables are mainly sensitive to the electroweakinos and sleptons, other irrelevant SUSY parameters, 
i.e., $A_{u,d,b,t,\ell}$, $M_3$, $M_{Q_L}$, $M_{U_R}$, $M_{D_R}$, $M_{E_3}$ and $M_{L_3}$, are set to $3\tev$ to be decoupled for simplicity. Note that one can set these parameters to other high mass scale but our conclusions are not changed. 

\begin{table}[b]
\begin{tabular}{|c|c|}
  \hline\hline
  Category & Experimental observables \\ \hline\hline
  DM relic density & $\Omega_\chi h^2= \Omega h^2 = 0.1186 \pm 0.002 \pm
  0.1\mu_t $~\cite{1502.01589} \\ \hline 
  $B$ physics &  
  ${\rm BR}(B \rightarrow X_s \gamma) = (3.27 \pm 0.14 \pm 0.1\mu_t) \times 10^{-4}$~\cite{1705.07933}  \\ 
  & ${\rm BR}(B^0_s \rightarrow \mu^+ \mu^-) = (3.0 \pm 0.6 \pm 0.3) \times 10^{-9}$~\cite{1703.05747}\\
  &  ${\rm BR}(B_u \rightarrow \tau \nu) = (1.09 \pm 0.24 \pm 0.1\mu_t) \times 10^{-4}$~\cite{Zyla:2020zbs}\\
   \hline
  Higgs physics & $R_{\rm{inv}} < 9\%$ at 95\% CL~\cite{ATLAS:2020qdt}\\ 
    & $m_{h_\mathrm{SM}} = (125.36 \pm 0.41 \pm 2.0)\gev$~\cite{1507.06706}\\  
   \hline
  DM DD & XENON1T~\cite{Aprile:2018dbl,1902.03234}, 
PICO-60~\cite{Amole:2019fdf}.\\    
  \hline\hline
 muon (g-2) &  $\gtwo^{\rm old} = (2.61 \pm 0.48 \pm 0.63) \times 10^{-9}$~\cite{Zyla:2020zbs} \\  
            &  $\gtwo^{\rm new} =  (2.51 \pm 0.59) \times 10^{-9}$~\cite{2104.03281} \\  

  \hline\hline
  GCE & As implemented in Ref.~\cite{Calore:2014nla}. \\
  \hline\hline
  LHC & $p p \rightarrow \chi_1^+ \chi_1^-$, $p p \rightarrow \chi_1^\pm \chi_2^0$ and $p p \rightarrow \tilde \ell_{L,R}^+ \tilde \ell_{L,R}^-$. \\   
  \hline\hline
\end{tabular}
\caption{The experimental constraints used in this study. 
}
\label{table:constraints}
\end{table}  

In Table~\ref{table:constraints}, we summarize the sets of experimental constraints 
that we invoke in the likelihood functions for the numerical scan. 
We define the total $\chi^2_\mathrm{tot}$ is the sum of $\chi_i^2$ where $i$ runs over 
all the constraints in table~\ref{table:constraints}. 
Beside the constraints (DM DD, GCE, and LHC), we use Gaussian likelihood for the rest constraints and 
their $\chi^2$ is defined as
\begin{equation}
    \chi^2 = \left( \frac{\mu_t - \mu_0}{\sigma} \right)^2~{\rm and}~
    \sigma = \sqrt{\sigma^2_\mathrm{theo} + \sigma^2_\mathrm{exp}}
\end{equation}
where $\mu_t$ is the theoretical prediction we calculated, $\mu_0$ is the experimental central value and 
$\sigma$ is the uncertainty including both theoretical and experimental errors.

We perform several random scans in the range as defined in Eq.~\eqref{eq:scan}.  
Except $\lambda$ and $\kappa$ are scan with log prior, the priors of 
the rest seven parameters are linear uniform distributed. 
We applied Metropolis-Hastings algorithm to undertake the focused scan at the parameter space 
with higher probability. 
The mass spectra and decay information are generated by using \texttt{NMSSMTools-5.5.2}~\cite{hep-ph/0508022}, 
and B-physics predictions are obtained by using \texttt{SuperIso-4.0}~\cite{0808.3144}. 
We use package \texttt{MicrOMEGAs-5.2.6}~\cite{1004.1092} for the calculation of DM relic density, 
muon $\gtwo$, and DM-nucleon cross sections.

Regarding the LHC constraints, we also consider the exclusions from the null results of 
searching for the SUSY events with two or three leptons plus missing transverse momentum at the 13 TeV LHC with the luminosities of 36.1 fb$^{-1}$~\cite{1803.02762} and 139 fb$^{-1}$~\cite{1908.08215}, respectively. 
We simulate the signal processes 
(i) $p p \rightarrow \chi_1^+ \chi_1^-$, 
(ii) $p p \rightarrow \chi_1^\pm \chi_2^0$ and 
(iii) $p p \rightarrow \tilde \ell_{L,R}^+ \tilde \ell_{L,R}^-$ are simulated by
\texttt{MadGraph5\_aMC-v3.1.0}~\cite{1405.0301} with default parton distribution function~\cite{1412.7420}. 
The next-to-leading order corrections to the cross sections of the above processes are included 
by using the factor $K=1.5$. 
Then the parton-level events are showered and hadronized with \texttt{PYTHIA-8.3}~\cite{1410.3012}. 
The detector effects are implemented by using \texttt{DELPHES-3.4.1}~\cite{1307.6346}. 
The package \texttt{CheckMATE-2.0.29} is used 
to recast the LHC analyses for each sample.
Finally, we define the event ratio $r = max(N_{S,i}/S^{95\%}_{obs,i})$ for each
experimental analysis, where $N_{S,i}$ is the number of the events for the $i$-th signal region 
and $S^{95\%}$ is the corresponding observed 95\% C.L. upper limit. 
The max is over all the signal regions for each analysis. 
As long as $r > 1$, we can conclude that such a sample is excluded at 95\% confidence limit (CL).

\section{Result}

\begin{figure}[!htbp]
    \centering
    \includegraphics[width=0.45\textwidth]{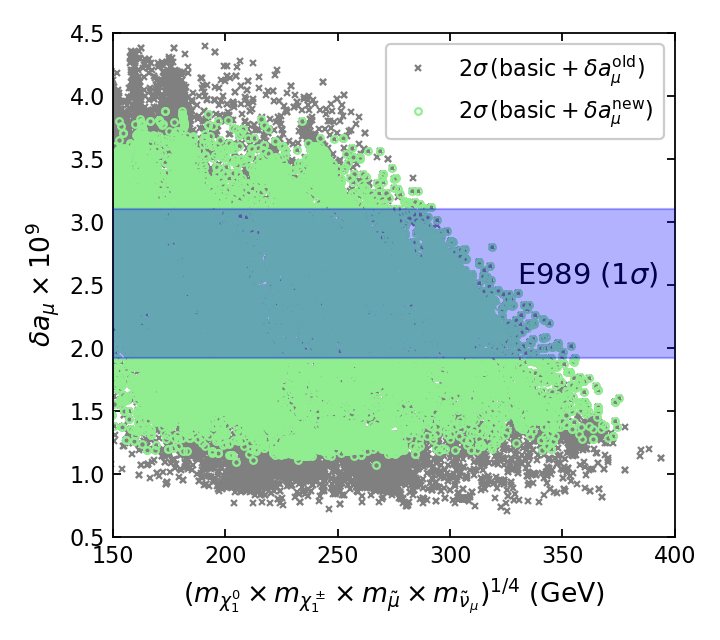}
    \caption{The distribution of $2\sigma$ allowed samples.  
     We predict the value of $\gtwo$ with respect to the geometric mean of the masses of 
     $\chi_1^0$, $\chi_1^\pm$, $\tilde \mu$, and $\tilde \nu_\mu$. 
    The grey (green) scatter points are $2\sigma$ allowed samples by  
    basic$+\gtwo^{\rm old(new)}$ constraints. 
    The shaded blue belt presents the E989 $1\sigma$ region.   
    }
    \label{fig:scale_vs_g2}
\end{figure}

To demonstrate the impact of new $\gtwo$ result and GCE signal on the NMSSM parameter space, 
we group the constraints except $\gtwo$, Fermi GCE, and LHC as basic set. 
Its statistic strength is denoted by $\chi^2(\mathrm{basic})$. 
We can see the role of new $\gtwo$ result by 
comparing the old and new $\gtwo$ result and 
we define their relevant chi-squares as $\chi^2(\mathrm{basic}+\gtwo^{\rm old})$ and $\chi^2(\mathrm{basic}+\gtwo^{\rm new})$. 
Finally, the Fermi GCE data squeeze the parameter space to close to $m_\chi\approx 60\gev$ 
and cross section around $\sv\approx 2\times 10^{-26}$~cm$^3 s^{-1}$. 
Our total chi-square including GCE data is $\chi^2(\mathrm{basic}+\gtwo^{\rm new}+{\rm GCE})$.  
We define the gray and green layers are $\delta\chi^2(\mathrm{basic}+\gtwo^{\rm old})<5.99$ (gray points), 
$\delta\chi^2(\mathrm{basic}+\gtwo^{\rm new})<5.99$ (green points), respectively. 
The top layer (red points) is with a slightly different definition. 
On top of the criteria $\delta\chi^2(\mathrm{basic}+\gtwo^{\rm new})<5.99$, 
we further require the survival red points to agree with LHC and GCE data within $95\%$ CL.

The propagator masses 
$m_{\chi_1^0}$, $m_{\chi_1^\pm}$, $m_{\tilde \mu}$ and $m_{\tilde \nu_\mu}$ 
enter the one-loop level of $\gtwo$ computation. In Fig.~\ref{fig:scale_vs_g2}, 
we show the correlation between the geometric average of these four masses and $\delta a_\mu$. 
The shaded blue belt is the $1\sigma$ error bar of new $g-2$ data. 
We find that the geometric average of these four masses has an upper limit at around $400\gev$ 
by applying $\gtwo^{\rm old}$ while the upper limit becomes $375\gev$ when updating to $\gtwo^{\rm new}$.

Since we are only interesting for a lighter neutralino mass region, 
the contribution of the neutralino-smuon loop with a light DM is usually dominant. 
As a drawback, the correlation between neutralino and smuon does not clearly appear 
from $\gtwo$ one-loop computation. 
We find from our scan that the production of mass scale $\sqrt{m_{\chi_1^\pm} m_{\tilde{\nu_\mu}}}$ 
is pushed to be less than $600\gev$ in $2\sigma$ after $\gtwo^{\rm new}$ is applied.  
Note that the contribution of second chargino $\chi_2^\pm$ can also be as important as 
the one from $\chi_1^\pm$ if they are nearly degenerated.

\begin{figure}[!htbp]
    \centering
    \includegraphics[width=0.48\textwidth]{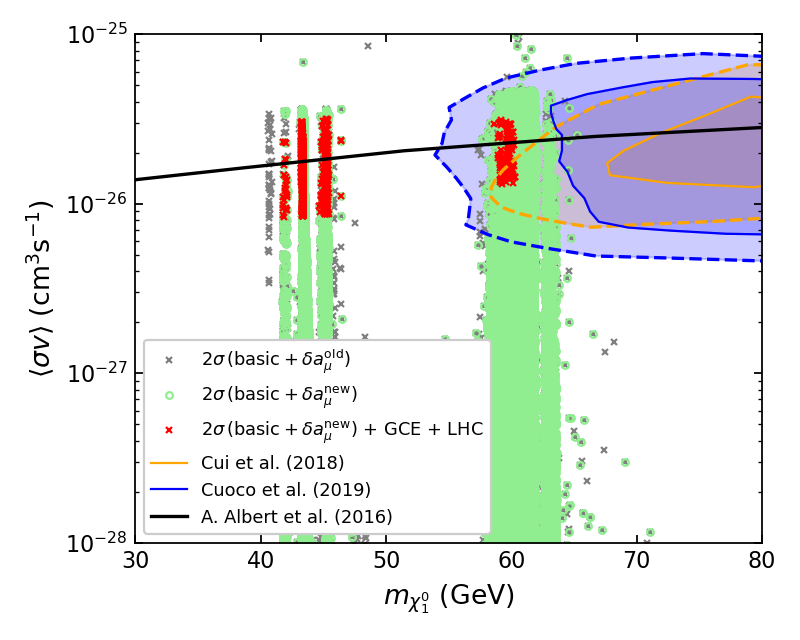}
    \caption{The annihilation cross section $\sv$ {\it {vs.}} $m_{\chi^0_1}$. 
    The grey, green and red dots are defined as legend. 
    The orange solid and dashed contours are for the anti-proton 68\% and 95\% C.L. from Ref.~\cite{1803.02163}. The blue solid and dashed contours are for the anti-proton 68\% and 95\% C.L. from Ref.~\cite{1903.01472}. Irrelevant region $\sv < 10^{-28} \mathrm{cm}^3 \mathrm{s}^{-1}$ was truncated. In this work, we do not include the bound set by the dwarf spheroidal galaxy observations~\cite{1611.03184} (black solid line), because some recent investigations show that the previous researches likely have significantly overestimated the stringentness of the limits~\cite{2002.11956, 2012.07846}.}
    \label{fig:sigmav}
\end{figure}

In Fig.~\ref{fig:sigmav}, we project all the samples of three groups on the ($m_{\chi^0_1}$, $\sv$) plane.  
It is clear that the allowed DM mass are around either near $Z$-resonance 
($m_\chi\simeq 45\gev$) or SM Higgs $H$ resonance ($m_\chi\simeq 60\gev$) 
in order to fulfill the relic density constraints. 
If anti-proton excess is also included, only the samples near 60 GeV survived. 
\textit{Therefore, we conclude that the bino-like DM in the NMSSM may be 
the common origin of muon $g-2$ anomaly, GCE and AMS-02 anti-proton excess.}

We would like to comment the possible constraints of the survived region 
near $60\gev$. 
Although the distortion of cosmic microwave background (CMB) power spectrum may be severe 
to DM annihilation to the leptonic final state, the CMB limits~\cite{Slatyer:2015jla} 
cannot exclude the survived region where $b\bar{b}$ is dominant the annihilation 
(more than $90\%$ of total contribution). 
We found the subdominant channel is $\chi\chi\to\tau^+ \tau^-$ which 
only contributes at most $10\%$ in total.
On the other hand, for $b\bar{b}$ final state annihilation, 
the Fermi gamma ray observations from dwarf spheroidal galaxies (dSphs)~\cite{Fermi-LAT:2016uux} can 
set a stronger bound than the one from CMB.  
In this work, we do not include the Fermi dSphs limit~\cite{ 1611.03184} but only present 
it in Fig.~\ref{fig:sigmav}, because some recent investigations show that 
the previous researches likely have significantly overestimated the stringentness of the limits~\cite{2002.11956, 2012.07846}.

To match the observed relic density, DM annihilation cross section in the early universe 
is around $10^{-26} \mathrm{cm}^3 \mathrm{s}^{-1}$. 
In our work, this can be achieved through SM-Higgs $h$ or $Z$ gauge boson resonance annihilation. 
Because the pure bino does not couple with $h$ or $Z$,   
the bino-like neutralino must contain some small fraction of higgsino ingredients 
in order to maintain $h$ or $Z$ resonance. 
If bino-like neutralino mass is not near the resonant area $\sim$ 45 GeV or $\sim$ 60 GeV, 
it would require a large composition of higgsino to reduce the relic density. 
However, we find that the current XENON1T data restricts the higgsino composition up to $\sim 2\%$.

Nevertheless, DM momentum in the present universe is no longer to maintain 
a large cross section via the $h$ and $Z$ resonance. 
Therefore, a new funnel is needed to undertake a cross section 
around $2\times 10^{-26} \mathrm{cm}^3 \mathrm{s}^{-1}$ for GCE. 
Unlike $CP$-conserving MSSM whose DM annihilation is $p-$wave suppressed at present,  
the pseudo-scalar mediator $A_1$ in the singlet sector of NMSSM 
can generate the $s-$wave process at zero temperature. 
One has to bear in mind that those annihilations in the present universe are via 
$A_1$-resonance even if it can be $Z$-resonance or $H$-resonance in the early universe.  
The final state of the $A_1$ funnel annihilation is governed by 
the $b\bar b$ channel with more than $90\%$ of total contribution. 
The subdominant annihilation final state is $\tau^+ \tau^-$ and 
it contributes at most $10\%$ in total.

\begin{figure}[!htbp]
    \centering
    \includegraphics[width=0.48\textwidth]{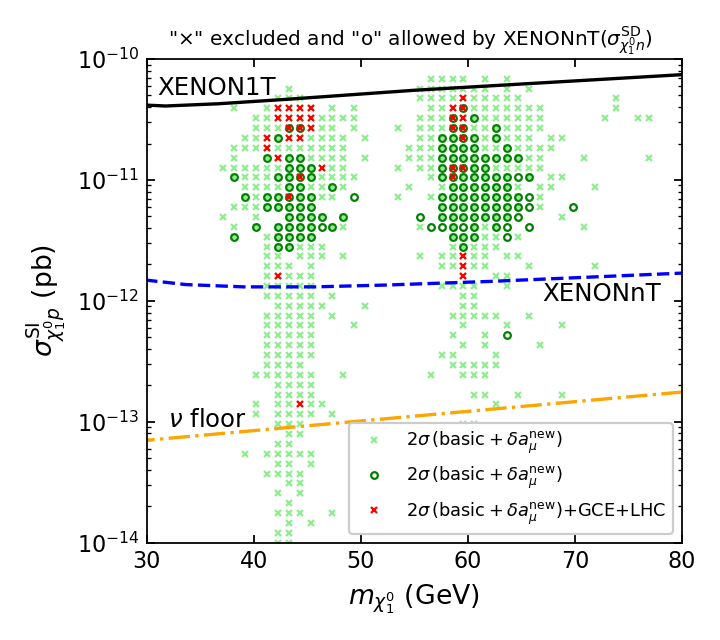}
    \caption{The DM-proton spin-independent cross section {\it {vs.}} DM mass $m_{\chi^0_1}$. 
    The dark and light green samples agree with $\chi^2(\mathrm{basic}+\gtwo^{\rm new})<5.99$.  
    If the green samples pass the $95\%$ limit of both GCE and LHC, 
    they are presented by the red color. 
    The black lines are XENON1T WIMP-proton SI~\cite{Gangi:2019zib} cross section $90\%$ upper limits. 
    The blue dashed lines are projected XENONnT~\cite{2007.08796} $90\%$ upper limits which are similar to the future PandaX-4T sensitivities \cite{PandaX:2018wtu}.
    The dot-dashed orange line is the neutrino floor.     
    The allowed samples by XENONnT $\sigsdn$ limit are marked by "$\circ$" 
    while those excluded samples are marked by "$\times$".
    \label{fig:DD}}
\end{figure}

In Fig.~\ref{fig:DD}, we plot the spin-independent (SI) component of 
the DM-proton elastic scattering cross sections in function of $m_{\chi^0_1}$. 
The black lines $90\%$ C.L. upper limits from XENON1T experiment~\cite{1902.03234, Gangi:2019zib}. 
There are some samples near the $\nu$ floor, namely the blind spot region~\cite{1705.09164}. 
Although $\sigsip$ in the blind spot region is very small and even unreachable below the neutrino floor, 
one can still probe them by future WIMP-neutron spin-dependent (SD) cross section $\sigsdn$ measurement 
as pointed out in Ref.~\cite{Banerjee:2016hsk}. 
We plot those samples allowed by XENONnT $\sigsdn$ limit~\cite{2007.08796} with "$\circ$" 
but those samples with $\sigsdn$ larger than XENONnT sensitivity are "$\times$"s. 
Indeed, the future XENONnT $\sigsdn$ sensitivity 
can probe those small $\sigsip$. 
Eventually, the parameter space favored by GCE can be completely probed by 
future XENONnT underground detector. 

Finally, we find that the current LHC SUSY particle searches  
are not able to completely exclude 
the region where all the excesses can be spontaneously explained. 
By scrutinizing the allowed charged particle masses with several $13\tev$ LHC analyses,  
we obtain the approximate upper limits $m_{\tilde \chi_1^\pm} \gtrsim 300\gev$ and $m_{\tilde \mu}\gtrsim 500\gev$.

\section{Conclusion}
In this Letter, we identify a common parameter space which can accommodate the muon $g-2$ anomaly, the GCE, and the anti-proton excess in the NMSSM. Considering various experimental constraints, e.g., the Higgs mass, B-physics, LHC data, DM relic density from PLANCK, and DM direct detections, 
we find that the light eletroweakinos and sleptons with masses being lighter than about 1 TeV are required. The geometric average of their masses should be less than about 375 GeV to explain the muon $g-2$ anomaly. Only the bino-like neutralino DM can explain both the muon $g-2$ and the GCE. They need to resonantly annihilate through $Z$ bosons or Higgs bosons to produce the correct relic density. On the other hand, in order to give enough DM annihilation cross section for the GCE, we need a singlet-like Higgs boson as the mediator in the $s$-channel resonance process at present. When further including the anti-proton excess, we find that only the Higgs funnel is feasible. The favored parameter space of the NMSSM model discussed in this work can be critically probed by the future XENONnT underground detector. 
We also expect that such a parameter space can be covered by 
the future AMS-02 as long as anti-deuteron or anti-Helium could be detected~\cite{Cholis:2020twh}.

\acknowledgments

This work is supported by the National Natural Science Foundation of China (NNSFC) under grant Nos. U1738210, 12047560, 11773075, by China Postdoctoral Science Foundation under grant No. 2020M681757, by Chinese Academy of Sciences, and by the Program for Innovative Talents and Entrepreneur in Jiangsu.

\bibliography{main}

\end{document}